\documentclass[12pt,onecolumn]{IEEEtran}
 \usepackage[colorlinks,bookmarksopen,bookmarksnumbered,citecolor=red,urlcolor=red,]{hyperref}
 \usepackage{indentfirst}
 \usepackage{algorithm,algorithmic,amsbsy,amsmath,amssymb,epsfig,bbm,mathrsfs, bbm} 
 \usepackage{multirow}
 \usepackage{mathrsfs}
 \usepackage{epstopdf}
 \usepackage[subfigure]{graphfig}
 \usepackage{bm}

\usepackage{hyperref}

\usepackage{xcolor}
 
 \usepackage{color}
 \usepackage{verbatim}
 \usepackage{amsfonts}
 \usepackage{amsthm}
\begin{document}
\title{Cyber Insurance for Heterogeneous Wireless Networks}

\author{Xiao Lu, Dusit Niyato, Hai Jiang, Ping Wang, and H. Vincent Poor
 }
\maketitle

\vspace{-16mm} 
 
\begin{abstract}
\vspace{-3mm}
Heterogeneous wireless networks (HWNs) composed of densely deployed base stations of different types with various radio access technologies have become a prevailing trend to accommodate ever-increasing traffic demand in enormous volume. Nowadays, users rely heavily on HWNs for ubiquitous network access that contains valuable and critical information such as financial transactions, e-health, and public safety. Cyber risks, representing one of the most significant threats to network security and reliability, are increasing in severity. To address this problem, this article introduces the concept of cyber insurance to transfer the cyber risk 
(i.e., service outage, as a consequence of cyber risks in HWNs) to a third party insurer. Firstly, a review of the enabling technologies for HWNs and their vulnerabilities to cyber risks is presented. Then, the fundamentals of cyber insurance are introduced, and subsequently, a cyber insurance framework for HWNs is presented. Finally, open issues are discussed and the challenges are highlighted for integrating cyber insurance as a service of next generation HWNs.

\end{abstract}
\vspace{-5mm}
\begin{IEEEkeywords} \vspace{-3mm}
Cyber insurance, cyber risk, heterogeneous wireless networks, risk assessment.
\end{IEEEkeywords} 
 
\vspace{-5mm} 
\section{Introduction}

Significant advances in information and communications technologies have resulted in increasingly more electronic devices of different use being connected to the Internet through pervasive wireless access. To provide seamless connectivity to these devices and accommodate their massive traffic demands, wireless networks are evolving towards a hierarchical architecture with denser deployment of different types of base stations (BSs) such as small cells and hotspots~\cite{H.2016Zhang}. 
Such hierarchical and dense networks are known as heterogeneous wireless networks (HWNs). 
HWNs support important wireless applications, for example, cloud-based services, e-business/commerce, e-health, and public safety applications. The loss, damage, or delay of data delivery in such applications can result in serious consequences.  
Thus, ensuring reliability and security of wireless data becomes one of the most important tasks in developing next generation HWNs.
 
Cyber risks in various forms of threats and attacks are major problems encountered by wireless communication and the services that it supports. Cyber risks become a major issue in network design, deployment, and operation to prevent cyber data exfiltration, denial-of-service (DoS), financial transaction compromise, and cyber extortion. Due to cyber risks, the financial loss incurred by the global economy is estimated to be 445 billion dollars in 2016 by Marsh $\&$ McLennan~\cite{MMC-Cyber-Handbook}.
With the ever increasing use of wireless networks of different types, damages caused by wireless attacks are envisioned to account for a considerable portion of the global financial loss caused by cyber risks.
The most common cyber risks in HWNs can be broadly sorted into active attacks, e.g., DoS attacks, and passive attacks, e.g., eavesdropping. The former impairs the {\em availability} while the later breaks the {\em confidentiality} of information on transmission.
With the boom of HWNs, users are progressively exposed to cyber risks that are consistently increasing in terms of diversity, number of occurrences and extent of the damage. For instance, amount of fraud loss on mobile devices and payment cards through contactless identity theft was reported to be 6.9 million British pounds in UK~\cite{FF_UK}.  
Addressing malicious attacks in HWNs features with the following difficulties 1) to pinpoint the attack sources because the trace can be disguised easily, 2) to find the attackers' identities due to anonymity of some Internet protocols, and 3) to avoid due to the open accessibility of wireless medium.

In view of the cyber risk damage, the global expenditure on cyber security is expected to reach 1 trillion from 2017 to 2021~\cite{RCD}. This potentially opens up a vast market for developing anti-risk solutions in HWNs. 
Nowadays, the mainstream of security research in HWNs focuses on developing system-based {\em risk mitigation} solutions in various enabling wireless technologies, such as massive Multiple-input Multiple-output (MIMO) technology, cloud radio access networks (C-RANs) and wireless caching~\cite{F.2016Akyildiz}. 
These enabling technologies aim to improve the robustness of wireless services and resilience against cyber risks.
However, there is a consensus in the research community that cyber security problems in HWNs cannot be completely eliminated by technological means alone~\cite{E.2014Hossain}. 
To tackle this problem, instead of mitigating the cyber risks, we take a different approach by introducing cyber insurance~\cite{L.2003A} as an alternative to shift the cyber risks in HWNs away from the operators and users. In particular, cyber insurance is a mechanism that transfers the risks undertaken by the operators or users of cyber systems to a third party, i.e., an insurance company. Cyber insurance differs from traditional insurance in two unique aspects: 1) cyber insurance needs to cover risks caused by smart and intentional attacks instead of natural failure; 2) cyber risks do not have geographical limitations, as simultaneous cyber risks can occur domain-wide, system/platform-wide, or even Internet-wide because of their identical or similar vulnerabilities. 
 
In HWNs, establishing a comprehensive cyber insurance framework is challenging. Firstly, the insurance policies for virtual products in cyber space (e.g., wireless applications and services) can rarely be derived from traditional insurance contracts due to the unique insurability of cyber insurance. 
The insurance coverage and liabilities must be re-defined. Secondly, risk assessment for the insurance buyer and insurer needs to be performed in the emerging wireless environments, with specific features such as network heterogeneity and randomness in geographical distribution, taken into investigation. New models need to be developed to characterize and analyze the risks for both the insurance buyer and insurer. These issues motivate us to design a novel cyber insurance framework for HWNs.

The main objectives of this article are: (i) to extend our understanding of traditional insurance to risk assessment in HWNs and outline research directions that may expand the use of cyber insurance in wireless environments, and (ii) to introduce novel analytical model to quantify important system and economic performance of the insurer. 
To this end, we introduce the fundamentals of cyber insurance in terms of concepts, terminologies, coverage, business models and procedures for forming a contract in HWNs. A cyber insurance framework is established for users in HWNs to relieve themselves from cyber risks. 
In particular, considering a general HWN in presence of a type of DoS attackers, i.e., wireless jammers, we take a quantitative approach to evaluate the vulnerability of a network user as the insured and the risks faced by the third-party insurer. 
Such an analytical investigation is crucial for formulating a cyber insurance contract in HWNs. 



The first novelty of this article is that we are the first to introduce a cyber insurance approach for a wireless environment to alleviate the damage from cyber risks. 
Different from existing cyber insurance approaches considered in wired networks, e.g., \cite{R.Zhang2017}, we focus on wireless jamming attacks in HWN environments, which can affect multiple users subject to stochastic fading channels and their spatial locations.   
Moreover, although some works analyzed the impact of network geographical distribution to the user's performance in HWNs, 
such as in \cite{H.2012Jo}, none has investigated the economic performance related to the cyber risk management. Another novelty is the introduction of a new analytical model that allows to investigate the impact of network and cyber attackers' geographical distributions on the economic performance of an insurer. The connection of network analysis and the economic analysis will open a new research direction to understand the interplay between wireless systems and cyber insurance business.


The rest of this article is organized as follows. Section~\ref{sec:overview} overviews the availability requirement, enabling technologies to improve availability and risk transfer in HWNs. Section~\ref{sec:cyberinsurance} provides the basics and fundamentals of cyber insurance. Section~\ref{sec:modeling} presents the models and analysis of the cyber insurance for HWNs. Section~\ref{sec:open} discusses open issues and research directions. Finally, Section~\ref{sec:conclusion} concludes the article.

\vspace{-3mm}
\section{Availability requirement, enabling technologies and risk transfer in Heterogeneous Wireless Networks}
\label{sec:overview}

\vspace{-2mm}
\subsection{Availability Requirement in HWNs}

One of the major security requirements in HWNs is {\em availability}. Availability is a characteristic of cyber security, which states that resources, services, and information within cyber systems are accessible in a timely and reliable manner. 
Availability becomes a pivotal factor of wireless systems as 
mobile data communications 
increasingly deliver critical information. An interruption in network service, or service unavailability, can lead to severe consequences especially for time-critical services including financial transactions, healthcare, system control, and public safety, etc. Therefore, it is imperative to provide seamless and timely data delivery between content service providers and residential users, businesses, and public sectors. 
\vspace{-5mm}
\subsection{Enabling Technologies  to Improve Availability in Heterogeneous Wireless Networks} 
In what follows, we introduce some major enabling technologies to improve availability in HWNs, namely, dense and heterogeneous deployment, massive MIMO, C-RAN and wireless caching. We highlight their benefits and limitations, with the aim of providing a better understanding of the underlying network structures and inherent characteristics of HWNs before presenting the discussion on cyber attacks to availability.


\subsubsection{Dense and heterogeneous deployment} 

A general trend in cellular networks is to complement macro-cells with multi-layered and dense deployment of small-cells with frequency reuse 
for enhanced coverage, higher data rate and decreased delay~\cite{H.2015Zhang}. 
Evidently, due to increased spatial reuse of network resources in HWNs, 
the impact of cyber attacks on service availability can be much more severe.

\subsubsection{Massive MIMO} 

Massive MIMO~\cite{F.2016Akyildiz}, or large-scale MIMO, adopts hundreds of antennas at BSs with the aim of pushing the spatial multiplexing, array gain, and degrees of freedom to the limit for better transmission capacity. 
However, cyber attacks, such as channel equalization attacks, can degrade the accuracy of channel estimation through pilot contamination so that the transmit
precoder design for massive MIMO is ineffective.

\subsubsection{Cloud Radio Access Networks}
The concept of C-RANs is to pool the baseband resources among multiple cell sites centralizedly  
so that they can be shared among the cell sites 
in an on-demand fashion. 
This architecture allows a flexible and dynamical resource distribution to handle nonuniform traffic so as to improve service availability especially in hotspots~\cite{F.2016Akyildiz}. 
Nevertheless, the centralized architecture of resource pool is subject to a single point of failure caused by DoS attacks.  

\subsubsection{Wireless Caching} 

Wireless caching or mobile content caching stores an extremely large amount of content 
from remote Internet servers
over the mobile networks in a distributed fashion, to locally serve
demand from end users. The locally available content can reduce the number of requests to the content publishers, thus 
alleviating delay and network congestion~\cite{E.Bastug2014}. 
Yet, cache devices are subject to DoS attacks that deteriorate the benefits of locally available content.

\vspace{-5mm} 
\subsection{Risk Transfer via Cyber Insurance}

As discussed above, all the reviewed enabling technologies for HWNs aim to improve the service availability. However, they also introduce vulnerabilities
that can be the targets of attacks to 
block availability. 
Again, it is apparent that technologies alone do not offer complete immunity from cyber risks. Thus, we take a different perspective by shifting the attention from risk mitigation to {\em risk transfer}. 
In particular, a cyber insurance framework is introduced to transfer the risks and damages in HWNs to a third-party insurer. 
In the past, cyber insurance as well as other traditional insurance options are available to typical computer and Internet users and businesses. 
For example, AIG~\cite{AIG} launched its CyberEdge PC policy in 2014 to offer cyber risk protection and recovery for property damage on a difference-in-condition basis.
However, these types of insurance including cyber insurance typically only cover losses of physical assets and known damage to cyber systems and user information. None of them extends the coverage to wireless environments. In the following, we introduce the concept of cyber insurance for HWNs, and propose a framework to assess the network performance of the users and economic performance of the insurer.

\vspace{-3mm}
\section{Cyber Insurance Fundamentals}
\label{sec:cyberinsurance}

This section reviews the fundamentals of cyber insurance, which serves as means to transfer cyber risks to a third party. 
\vspace{-5mm}
\subsection{Concepts and Terminology}

\begin{figure}
	\centering
	\includegraphics[width=0.9\textwidth]{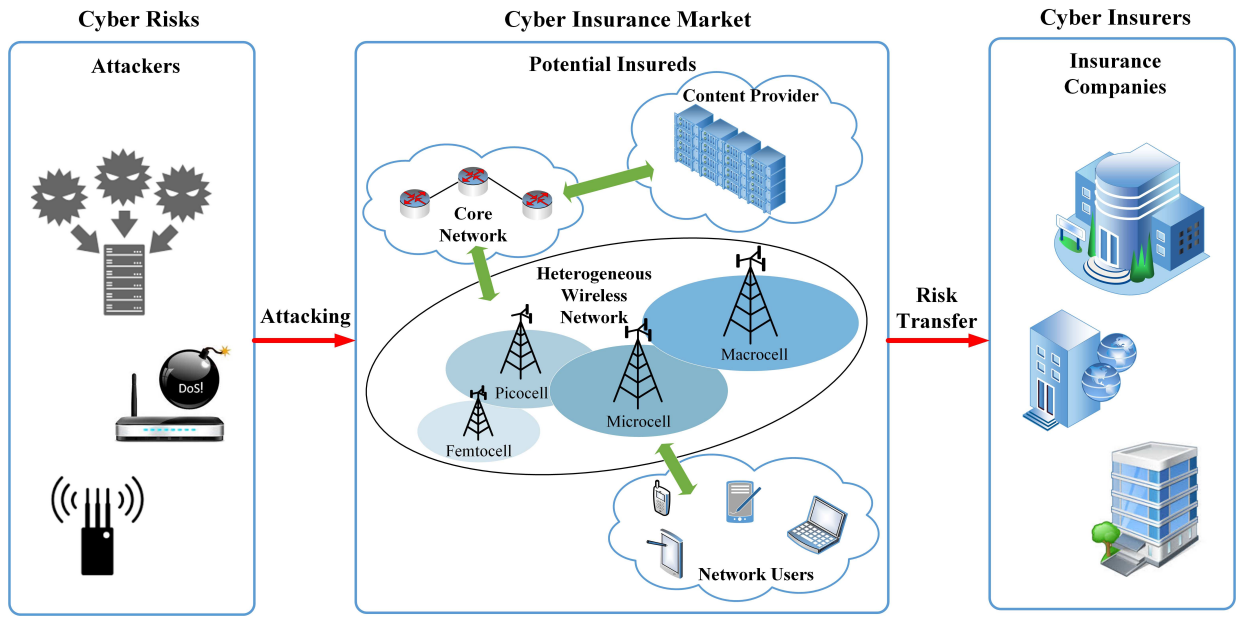} 
	\caption{Business models for cyber insurance markets} \label{fig:business}
\end{figure}

As shown in Fig.~\ref{fig:business}, cyber insurance is a risk-transfer mechanism between a party exposed to potential cyber risks and another party that is willing to compensate the losses and damages resulting from such risks. For the former which is referred to as an insured, cyber insurance is to provide financial compensation in case of loss at a fixed cost to be insured. For the latter party which is referred to as an insurer, it affords liabilities resulting from the uncertain risks in the future in exchange for certain monetary benefit from the former party in advance. We introduce the terminology commonly used in cyber insurance as follows: 

\begin{itemize}

 
\item {\bf Cyber risks} refer to exposure to any potential threats, not limited to cyber attacks, that could cause 
damage to the information and technology assets. 

 

\item A {\bf Cyber insurance contract} is an agreement between the cyber insurer and the cyber insured that specifies the amount of payment by the insured and the liabilities of the insurer.
 
\item A {\bf Premium} is the amount of payment charged by an insurer for a cyber insurance contract.
 
\item A {\bf Claim} is a formal request to a cyber insurer activated by a cyber insured for compensation regarding the occurrence of losses.

\item An {\bf Indemnity} is a financial compensation for the losses or damages originated by cyber attacks.

\end{itemize}

In general, the losses resulting from cyber attacks can be {\em primary losses} or {\em secondary losses}. The former refer to the direct and initial costs, and the latter comprises the losses that follow as a consequence of the former. For example, the amount of data that the users cannot transfer due to an HWN disruption can be considered to be a primary loss, whereas the users that 
switch to other competing network service providers and reputation damage of the corresponding services can be considered to be secondary losses.

\vspace{-5mm}
\subsection{Cyber Insurance Procedure for Network Users}

Issuing a cyber insurance contract comprises five major steps. 

\begin{enumerate}
	\item {\bf Risk Identification:} The entity to be insured first determines the potential risks that it is facing and identifies the risks from which it wants to be protected. 

	\item {\bf Insurer Selection:} The entity to be insured analyzes the available insurers on the market and initiates an insurance proposal/request to the chosen insurer. 

	\item {\bf Customer Evaluation:} Upon receiving the proposal, the insurer collects the required information from the customer to evaluate its potential cyber risks. 
 
	\item {\bf Contract Establishment:} Based on the evaluation, the insurer offers a cyber insurance contract with specified coverage and amount of premium to the customer. If the customer accepts and signs the offer with the insurer, the contract becomes valid and the cyber insurance procedure moves to the next step. Otherwise, the customer can negotiate with the insurer the details in the contract, e.g., premium, coverage, and indemnity, until reaching a 
	mutually agreeable contract, or decline the contract if no common agreement can be reached. 

	\item {\bf Contract Execution:} During the effective period of the contract, the insurer undertakes the cyber risks of the insured according to the agreement. If losses occur, the insured has the right to make a claim and receive the corresponding indemnity from the insurer after the claim is verified. 
 \end{enumerate}

\subsection{Risk Assessment}

Risk assessment that determines the risk exposure of an insured is an essential task for an insurer. A cyber insurance contract can be established only if the insurer has appropriate awareness of cyber risks that it will undertake. 
However, a comprehensive risk assessment is challenging and resource-consuming in two aspects: Firstly, it demands information acquisition from different parties, e.g., the insureds, the cyber system operators and information technology consulting firms. Secondly, assessment models need to be developed to acquire quantitative estimates of cyber risks. 
The results of risk assessment provide guidelines to associate each risk with a price, which helps to provide a basis for premium and indemnity setting. 
These risk quantifications may also serve as incentives for the insureds to take risk-appropriate behavior and risk mitigation approaches to reduce their premiums. Yet, such a risk assessment model for HWNs is missing in the existing literature, which is the focus of the next section.



\vspace{-3mm}
\section{Modeling and Performance Analysis of Cyber Insurance for HWNs}
\label{sec:modeling}

In this section, we introduce a cyber insurance framework for HWNs. 
We consider a dense HWN in presence of randomly located DoS attackers, i.e., wireless jammers, that broadcast noise to deteriorate the service availability.  
In this HWN, both wireless attacks (e.g., jamming) and inefficient resource allocation (e.g., heavy frequency reuse) 
can cause excessive interference to impair the users' wireless access and even cause potential service outage.
A cyber insurance framework is introduced for the network users as the insured to be protected from such breaches and for the insurer to evaluate its own benefits.  
In this framework, the users face network risks that may result in service outage. Correspondingly, the insurer carries financial risks that it may not have sufficient money to offset the claims from the users, referred to as a {\em ruin}. Thus, we consider 
{\em service outage probability} and {\em ruin probability} as the risk metrics for the users and insurer,
respectively, which will be defined in the following subsections. 
The insurer needs to evaluate the average service outage probability of a generic user in a large-scale HWN. Based on this service outage probability, the insurer can analyze its own ruin probability by evaluating the amount of premium and the amount of indemnity 
in case of service outage. 

In the following subsections, we first describe the network model under consideration, present the cyber insurance framework, and then demonstrate the performance analysis of the network and framework. 
\vspace{-5mm}
\subsection{Network Model of Heterogeneous Wireless Networks}
\label{sec:Model}

\begin{figure}
	\centering
	\includegraphics[width=0.65 \textwidth]{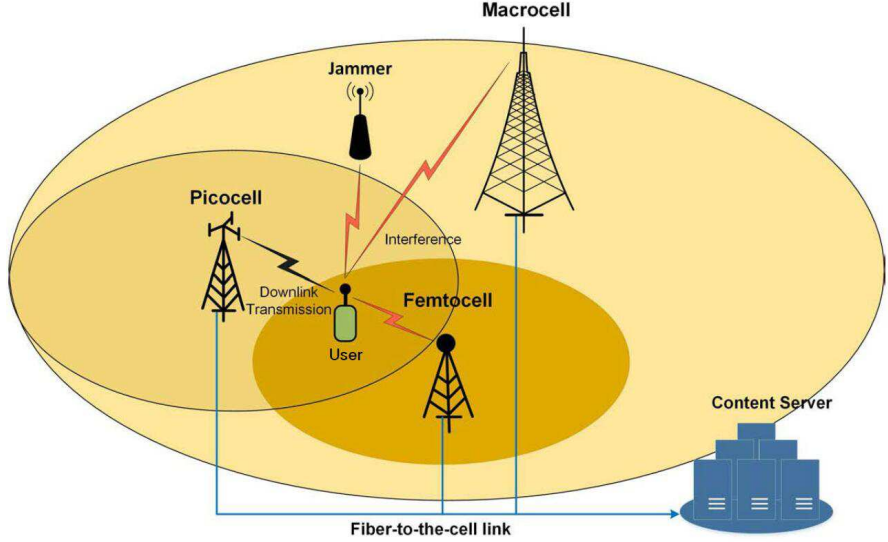} \vspace{-6mm}
	\caption{Illustration of a $K$-tier ($K=3$) heterogeneous wireless network.} \label{fig:illustration}
	\vspace{-3mm}
\end{figure}

We consider wireless jamming as the cyber attack in HWNs. 
as shown in Fig.~\ref{fig:illustration}. The service provider provisions downlink wireless service to the users through a $K$-tier architecture in the presence of malicious jammers. The $K$ network tiers are composed of BSs using different transmit powers $P_{k}$, spatial densities $\zeta_{k}$, distribution repulsion degrees $\alpha_{k}$, and path loss exponents $\mu_{k}$, $k \in \mathcal{K}$ here $\mathcal{K}=\{1,2,\cdots,K\}$ denotes the set of the network tiers. We model the distribution of the BSs of the $K$-tier HWN as $K$ independent homogeneous point processes $\Phi_{k}$. 
Each tier can be characterized as $\{\Phi_{k}, P_{k}, \alpha_{k},\zeta_{k}, \mu_{k} \}, \forall k \in \mathcal{K}$. Additionally, the malicious jammers are considered to be radio-frequency transmitters that broadcast noise to damage the downlink transmissions of the HWN. The jammers are randomly distributed modeled by a homogeneous point process denoted as $\Phi_{J} $. 
Similarly, the set of jammers is characterized as $\{\Phi_{J}, P_{J}, \alpha_{J},\zeta_{J}, \mu_{J} \}, \forall k \in \mathcal{K}$, where $P_{J}$, $\alpha_{J}$, $\zeta_{J}$, $\mu_{J}$ denote the transmit power, repulsion degrees, spatial density and path loss exponent for the jammers, respectively. In this article, we characterize $\Phi_{k}$ and $\Phi_{J}$ based on a general framework, namely, using a $\alpha$-Ginibre point process (GPP)~\cite{L.2015Decreusefond}. 

We assume that all the BSs are connected to the core network via fiber-to-the-cell links. An open network access policy is used so that the wireless service can be provided to the subscribers or users by any network tier through downlink transmission. For cell association, a maximum received power scheme is adopted~\cite{H.2012Jo}, where a user is served by a BS that provides the strongest average received power. Since the users have a random geographical distribution, the number of users connected to each BS is random. The group of users served by the same BS are allocated with orthogonal frequency resource blocks, i.e., there is no intra-cell interference. Among the BSs in the same network tier, we consider frequency reuse with a factor $\xi \in (0,1]$ indicating the percentage of BSs operating on the same spectrum frequency. Furthermore, we consider an underlay heterogeneous network, in which all the network tiers share the same frequency resource pool. 
As for the wireless channels, we consider path loss attenuation plus block Rayleigh fading.
An example of a three-tier HWN (i.e., Macrocell, Picrocell and Femtocell) is depicted in Fig~\ref{fig:illustration}. In this example, the user is associated with the picocell tier. Due to frequency reuse, the downlink transmission may be subject to interference from the Macrocell and Femtocell tiers. Additionally, the noise from the malicious jammers also deteriorates the information signal received at the user.

We define a service outage as an event that the received signal-to-interference-plus-noise ratio (SINR) at a user from its associated BS is less than a predefined threshold $\tau$~\cite{H.2012Jo}.
Based on the above $K$-tier HWN model and $\alpha$-GPP modeling, 
we characterize the service outage probability of a generic user 
in the HWN. Due to space limitations, we omit the derivation. 
\vspace{-5mm}
\subsection{Cyber Insurance Framework}

We propose a cyber insurance framework in which an insurer offers a cyber insurance contract to users of an HWN. 
First, each mobile user pays the premium 
to the insurer. In the event that service outage occurs to the user, e.g., because of interference or a jamming attack, the insurer compensates the user with a claim amount. 
For the insurer, its income is from the premium paid by $U$ users. On the other hand, the expense of the insurer is the claims generated randomly from the user population given a service outage probability. The insurer is, therefore, interested in its ruin which corresponds to the event that its reserve, i.e., initial balance and accumulated income minus aggregated indemnity, is negative. The ruin is important to the insurer as it indicates the state in which the insurer does not have enough balance to pay for incoming claims. This can cause a breach of contract, harming the insurer's business. 

In the following, we analyze the ruin probability of the insurer~\cite{dickson2010}. We model the number of claims $N_{t}$ until time $t \ge 0$ by a homogeneous Poisson process $(N_t)$ with intensity $\lambda >0$. The intensity $\lambda$ 
is the product of the service outage probability derived in Section~\ref{sec:Model} and the number of users $U$. 
Let $(W_k)$ denote a sequence of independent non-negative, identically-distributed random claim amounts, $Y_k$ 
denote the aggregate amount of $k$ claims, and $f(t)$ denote a non-decreasing, time-dependent function of the premium income 
recevied until time $t$. According to the compound Poisson risk model, the summation of claim amount over the considered time duration can be expressed by $Y_{N_t}$. 
Let $y \geq 0$ denote the amount of initial reserve. The surplus process of the insurer's reserve 
is defined as the initial reserve plus the premium income with the claim amounts subtracted, 
i.e., $R(t) =y+f(t) -Y_{N_{t}}$. 

We consider a finite time horizon until moment $T$. 
Then, the finite-time ruin probability can be characterized as the probability that $R(t)$ falls below zero at a time instant before moment $T$. 
The insurer is interested in the lowest level of its reserve, denoted as ${\mathcal M}_{[0,T]}$, 
between time $0$ and some fixed time horizon $T>0$. This is an explicit probabilistic representation for a compound Poisson process. It corresponds to the classical Cr\'amer-Lundberg risk model which is suitable for simulation purposes. In the following, we 
refer to~\cite{piclef} for a direct integration by parts on the Poisson space to compute the density of the infimum ${\mathcal M}_{[0,T]}$. 
These performance metrics can be obtained efficiently using Monte Carlo methods.

\vspace{-3mm}
\subsection{Numerical Results}

\begin{figure}
	\centering
	\includegraphics[width=0.45 \textwidth]{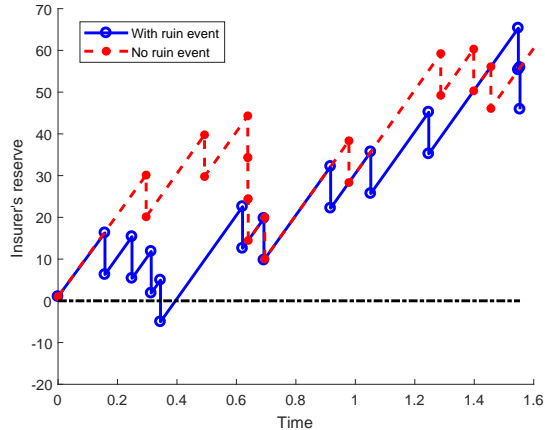} 
	\caption{Samples of insurer's reserve.} \label{fig:sample} 
\end{figure}

In this subsection, we evaluate 
the cyber insurance framework in an HWN. We consider a two-tier HWN coexisting with randomly distributed malicious jammers. There are 1,000 network users buying insurance contracts. Unless otherwise stated, the premium is 0.1 monetary units (MUs), the initial reserve of the insurer is 1.0 MU, and the insurance period is 10 time units. We set the transmit powers $P_1$, $P_2$ and $P_J$ as 40, 33 and 30 dBm, respectively. The path loss exponents $\mu_1$, $\mu_2$, and $\mu_J$ are 3.5, 4, and 4, respectively. The density of the tier-one BS is $\zeta_{1}=0.002$.

We first show an example of the insurer's reserve over time. 
With the service outage probability 0.01, 
Fig.~\ref{fig:sample} illustrates the insurer's reserve with and without ruin, respectively. In both cases, the reserve increases linearly due to the constant premium rate. However, when an outage happens to one of the users, the insurer has to pay the claim, and as such, the reserve sharply drops. A ruin event happens when such a drop makes the reserve fall below zero as shown in the blue plot. When a ruin event happens, the insurer does not have enough reserve to pay for the claim. Therefore, it is important to analyze the ruin probability which determines the insurer's vulnerability to insolvency.

\begin{figure} 
\centering
 \subfigure [The service outage probability of a generic user as a function of the density of jammers $\zeta_J$.] {\label{fig:Outage_probability}
 \centering
 \includegraphics[width=0.45 \textwidth]{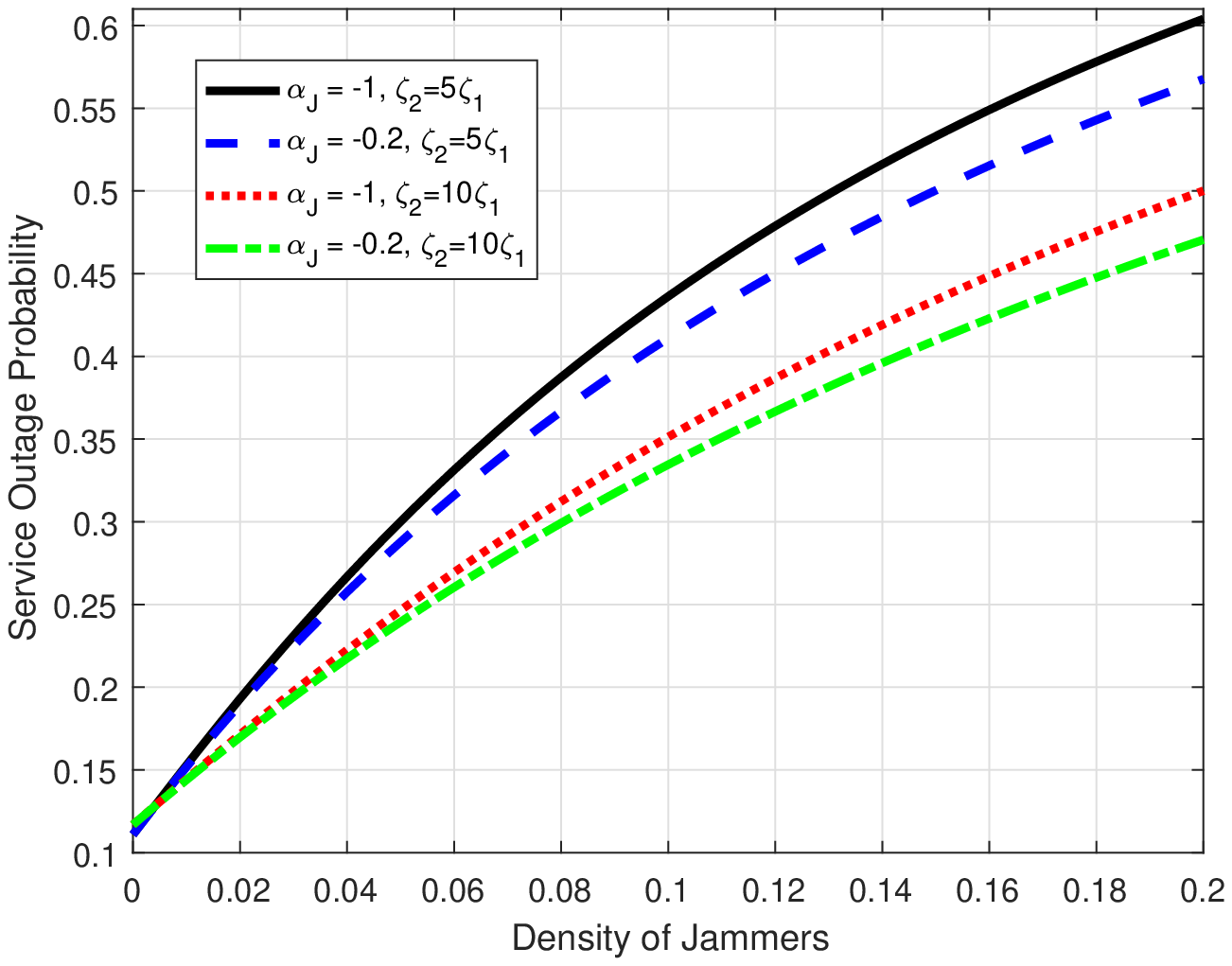}} 
 \centering
 \subfigure [The ruin probability of the insurer as a function of the density of jammers $\zeta_{J}$] {
 \label{fig:RP_densityJ} 
 \centering
\includegraphics[width=0.45 \textwidth]{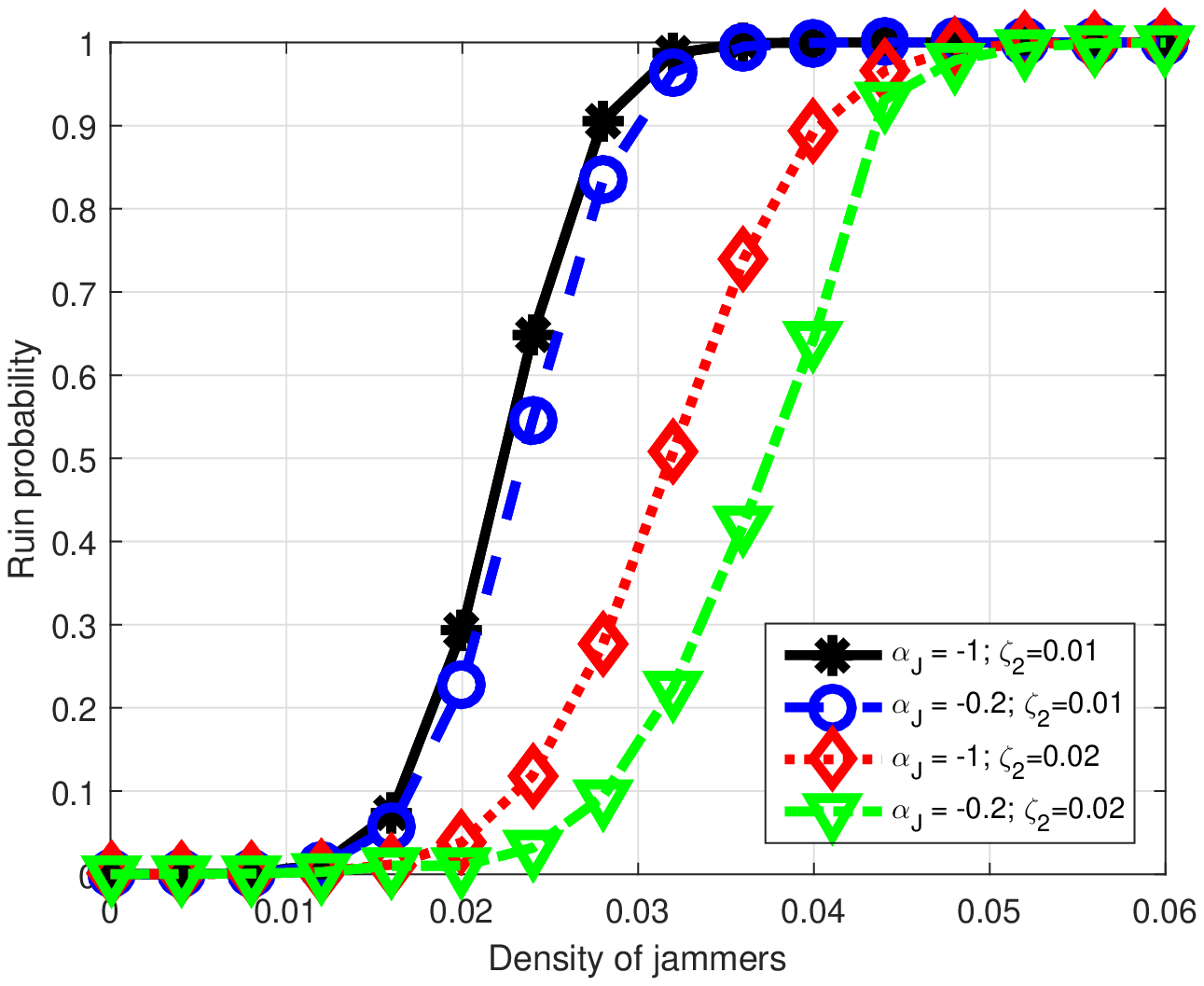}}\\
\caption{The Impact of Interference Signals (Frequency reuse factor $\xi=1$).}
\centering
\label{fig:throughput}
\vspace{-3mm}
\end{figure}

Next, we investigate the impact of jamming signals on the insureds and insurer. 
Figure~\ref{fig:Outage_probability} demonstrates the effect of repulsion (i.e., $\alpha_{J}$) and density (i.e., $\zeta_{J}$) of malicious jammers on service outage probability when the SINR threshold $\tau$ is -20 dB. We observe that 
when $\zeta_{J}$ increases, the service outage probability increases fast in small $\zeta_{J}$ region and slowly in large $\zeta_{J}$ region.
Moreover, a stronger repulsion among the jammers leads to a larger service outage probability. 
The reason is that stronger repulsion among the jammers generates more interference to the users, causing larger performance
degradation. Next, in Fig. \ref{fig:RP_densityJ}, we show the ruin probability of the insurer as a function of the density of jammers. In comparison with the service outage probability in Fig.~\ref{fig:Outage_probability}, it can be seen that a larger density of jammers and/or a stronger repulsion among jammers also lead to a larger ruin probability.  
Moreover, the ruin probability
increases with the density of jammers with a much faster rate than the service outage probability. 

\begin{figure}
	\centering
	\includegraphics[width=0.45 \textwidth]{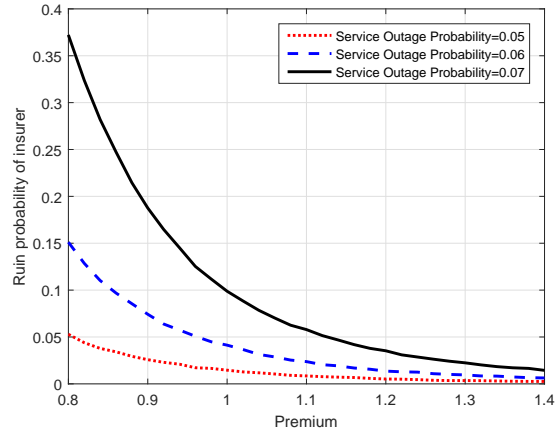} \vspace{-3mm}
	\caption{The ruin probability as a function of premium rate.} \label{fig:ruin_premium}
\end{figure}

Next, we study the risk of the insurer. Fig.~\ref{fig:ruin_premium} shows the ruin probability as a function of premium rate under different values of the service outage probability. We can see that the ruin probability is a monotonically decreasing function of the premium rate. 
It is worth noting that when the premium rate is small, the ruin probability is very sensitive to the service outage probability. When the premium is low (e.g., 0.8), a small increase of service outage probability can result in a significant increase of ruin probability. Thus, in order to manage its own ruin probability, the insurer needs to set the premium based on the user's service outage probability.


\section{Open Issues and Future Directions}
\label{sec:open}
This section discusses some open issues in cyber insurance research and sheds light on possible future directions.


\begin{itemize}

\item {\em Premium and Indemnity Calculation:} Premium and indemnity affect the profit of the insurer and satisfaction of users in HWNs. Their optimization is required to balance the tradeoff between the users' satisfaction with cyber insurance and the ruin probability of the insurer. The proposed analysis can serve as a tool for such optimization.


\item {\em Cyber Risk Correlation:} Different from the classical risks, e.g., car crashes and theft, which have geographical independence, 
cyber risks in HWNs involve attacks on information and cyber systems that may exhibit correlation. For example, an attacker who performs DoS attacks may also eavesdrop the confidential information of a user.
Therefore, the risk correlation significantly affects insurers' decisions in the premium setting to avoid a ruin event. The measure and quantification of risk correlation among different losses impose significant challenges in finding optimal premiums.

\item  {\em Insurability:}  As yet, the insurability of emerging wireless and mobile services needs to be determined and quantified for which novel mathematical and statistical models are needed. 
For example, in the case of harmful jamming, a mechanism is needed to quantify the damage of the user.


\item {\em Compensation of Secondary Loss:}  The calculation of the indemnity for a cyber insurance claim can be complicated if secondary losses are involved. 
An instance is that after cyber risks generate the primary loss to a service provider's network infrastructure, the typical secondary loss following is that users change their subscriptions 
to other service providers due to dissatisfaction caused by the cyber risks. 
Such a secondary loss may be hard to observe and take a long time before causing an effect, which needs to be explored in establishing a mature cyber insurance contract. 

\end{itemize}

\vspace{-3mm}
\section{Conclusion}
\label{sec:conclusion}

In this article, we have first discussed the limitations of the enabling technologies for HWNs in defending against cyber attacks. Then, we have presented a cyber insurance framework for next generation HWNs to transfer to the insurer losses and damages caused by service outages, e.g., due to cyber attacks and system failures. 
For a large-scale $K$-tier HWN, we have characterized the user risk performance in terms of service outage probability. Then, we have introduced analysis of the ruin probability for the insurer. The analysis will be useful for  setting the premiums and indemnities to achieve a better tradeoff between the benefits of the insurer and/or users.

\begin{IEEEbiography}  \! 
\textbf{ Xiao Lu} is now pursing the Ph.D degree in the University of Alberta, Canada. He received an M.Eng.
degree in computer engineering from Nanyang Technological University, and a B.Eng. degree in communication
engineering from Beijing University of
Posts and Telecommunications. His current research
interests are in the area of stochastic modeling and
analysis of wireless communications systems.
\end{IEEEbiography} 

\begin{IEEEbiography}  \! 
\textbf{Dusit Niyato} (M’09-SM’15-F’17) received the
B.Eng. degree from the King Mongkuts Institute of
Technology Ladkrabang, Thailand, in 1999, and the
Ph.D. degree in electrical and computer engineering
from the University of Manitoba, Canada, in
2008. He is currently an Associate Professor with
the School of Computer Science and Engineering,
Nanyang Technological University, Singapore. His
research interests are in the area of energy harvesting for wireless communication, Internet of Things, and sensor networks.
\end{IEEEbiography} 
 
\begin{IEEEbiography}  \!
\textbf{ Hai Jiang} (SM’15) received the B.Sc. and M.Sc. degrees in electronics engineering from Peking University, Beijing, China, in 1995 and 1998, respectively, and the Ph.D. degree in electrical engineering from the University of Waterloo, Waterloo, Ontario, Canada, in 2006. He is currently a Professor at the Department of Electrical and Computer Engineering, University of Alberta, Canada. His research interests include radio resource management, cognitive radio networking, and cooperative communications.
\end{IEEEbiography}

\begin{IEEEbiography}\!
\textbf{Ping Wang} (M’08-SM’15) received the Ph.D. degree in electrical engineering from the University of Waterloo, Canada, in 2008. She is currently an Associate Professor with the School of Computer Science and Engineering, Nanyang Technological
University, Singapore. Her current research interests include resource allocation in wireless networks, cloud computing, and smart grid. She was a corecipient of the Best Paper Award from the IEEE Wireless Communications and Networking Conference in 2012 and the IEEE International Conference
on Communications in 2007.
\end{IEEEbiography}

\begin{IEEEbiography}\! 
\textbf{H. Vincent Poor} (S’72-M’77-SM’82-F’87)
received the Ph.D. degree in EECS from Princeton
University in 1977. 
Since 1990, he has been
on the faculty at Princeton, where he is currently
the Michael Henry Strater University Professor
of Electrical Engineering. From 2006 to 2016,
he served as the Dean of Princeton’s School of
Engineering and Applied Science. His research
interests are in the areas of information theory,
statistical signal processing and stochastic analysis, and their applications in wireless networks and related fields such as smart grid and social networks.   
\end{IEEEbiography} 
 
\end{document}